\documentclass[12pt]{article}
\usepackage[margin=.75in]{geometry}
\usepackage{amsmath, amssymb, amsthm}
\usepackage{mathtools}
\usepackage{graphicx}
\usepackage{array}
\usepackage{amsfonts}

\pdfoutput=1
\usepackage{putex}
\usepackage[vcentermath]{youngtab}
\usepackage{lscape}
\usepackage{cite}
\usepackage{graphicx}
\usepackage{epstopdf}
\usepackage{enumerate}
\usepackage{tensor}
\usepackage{slashed}
\usepackage{amsmath}
\usepackage{amssymb}
\usepackage{mathrsfs}
\usepackage{lgrind}
\usepackage{amsmath,amssymb,braket}
\usepackage{bbm}
\usepackage{hyperref}
\usepackage{subcaption}

\usepackage{tikz}

\numberwithin{equation}{section}

\newcommand {\be} {\begin {equation}}
\newcommand {\ee} {\end {equation}}

\newcommand {\bes} {\begin {equation*}}
\newcommand {\ees} {\end {equation*}}

\newcommand{\eps}{\epsilon}

\def\CH{{\cal H}}

\newcommand{\beq}{\begin{equation}}
\newcommand{\eeq}{\end{equation}}

\def\be{ \begin{equation} }
\def\ee{ \end{equation} }

\def \be {\beta}

\def \beq { \begin{equation}}
\def \eeq {\end{equation}}

\renewcommand{\tr}{\operatorname{tr}}

\usepackage{indentfirst}
\begin{document}
\preprint{PUPT-2610}
	
\institution{CU}{St Edmund's College, University of Cambridge,
	Cambridge, CB3 0BN,	UK}
\institution{PU}{Department of Physics, Princeton University, Princeton, NJ 08544, USA}
\institution{PCTS}{Princeton Center for Theoretical Science, Princeton University, Princeton, NJ 08544, USA}

\authors{G. Gaitan,\worksat{\CU} I.R. Klebanov,\worksat{\PU,\PCTS} K. Pakrouski,\worksat{\PU} P.N. Pallegar,\worksat{\PU}  F.K. Popov\worksat{\PU} }
\title{Hagedorn Temperature in Large $N$ Majorana Quantum Mechanics }

\abstract{We discuss two types of quantum mechanical models that couple large numbers of Majorana fermions and have orthogonal symmetry groups. In models
of vector type, only one of the symmetry groups has a large rank. The large $N$ limit is taken keeping $gN=\lambda$ fixed, where $g$ multiplies the quartic Hamiltonian. We introduce a simple model with $O(N)\times SO(4)$ symmetry, whose energies are expressed in terms of the quadratic Casimirs of the symmetry groups. This model may be deformed so that the symmetry is $O(N)\times O(2)^2$, and the Hamiltonian reduces to that studied in \cite{Klebanov:2018nfp}. We find analytic expressions for the large $N$ density of states and free energy.  
In both vector models, the large $N$ density of states varies approximately as
$e^{-|E|/\lambda}$ for a wide range of energies. This gives rise to critical behavior as the temperature approaches the Hagedorn temperature $T_{\rm H} = \lambda$. 
In the formal large $N$ limit, the specific heat blows up as $(T_H- T)^{-2}$, which implies that $T_H$ is the limiting temperature. However, at any finite $N$, it is possible to reach arbitrarily large temperatures.
Thus, the finite $N$ effects smooth out the Hagedorn transition.
We also study models of matrix type, which have two $O(N)$ symmetry groups with large rank. An example is provided by the Majorana matrix model with $O(N)^2\times O(2)$ symmetry, which was studied in \cite{Klebanov:2018nfp}. In contrast with the vector models, the density of states is smooth and nearly Gaussian near the middle of the spectrum.
}
	
\date{}
\maketitle
\tableofcontents

\section{Introduction and summary}

Strongly interacting fermionic systems describe some of the most challenging and interesting problems in physics. For example, one of the big open questions in condensed matter physics is the microscopic description of the various phases observed in the high-temperature superconducting materials. Models relevant in this context~\cite{ReviewVariousOrdersOfHighTc,WenReview2006tJandHubbardvsCuprates,GeorgesDMFTHubbard} include the Hubbard~\cite{HubbardHubbard63,veryHighLevelHubbardReview} and t-J models \cite{SPATEKtJ77}.  The Hamiltonians of these models include the quadratic hopping terms for fermions on a lattice, as well as approximately local quartic interaction terms.
The analysis of such models often begins with treating a quartic interaction term as a small perturbation. In the cases when such an expansion is not possible, 
for example, the fractional quantum Hall effect, one typically has to resort to numerical calculations. 

Fortunately, there are also fermionic systems which can be solved analytically in the strongly interacting regime, when the number of degrees of freedom is sent to infinity. Such large $N$ systems include the Sachdev-Ye-Kitaev (SYK) models \cite{Sachdev:1992fk,Kitaev:2015,Sachdev:2015efa,Polchinski:2016xgd,Maldacena:2016hyu,Kitaev:2017awl} (see also the earlier work \cite{Bohigas:1971vpj,French:1971et}). The SYK models have been studied extensively in the recent years; for reviews and recent progress, see \cite{Sarosi:2017ykf,Rosenhaus:2019mfr,Gu:2019jub}. 

The simplest of them, the so-called Majorana SYK model \cite{Kitaev:2015,Kitaev:2017awl}, has the Hamiltonian $H = 
J_{ijkl} \psi^i \psi^j \psi^k \psi^l$, which describes a large number $N_{\rm SYK}$ of Majorana fermions $\psi^i$ (we assume summation over repeated indices throughout this work). They have random quartic couplings $J_{ijkl}$ with appropriately chosen variance. A remarkable feature of this model is that, in the limit where 
$N_{\rm SYK}\rightarrow \infty$, it becomes nearly conformal at low energies. The low-lying spectrum exhibits gaps which are exponentially small in $N_{\rm SYK}$. 
In further work, models consisting of coupled pairs of Majorana SYK models \cite{Maldacena:2018lmt,Garcia-Garcia:2019poj,Kim:2019upg}, as well as the SYK chain models \cite{Gu:2016oyy,Patel:2018zpy}, have produced a host of dynamical phenomena which include gapped phases and spontaneous symmetry breaking. In addition to the terms quartic in fermions, they can include quadratic terms which describe hopping between different SYK sites.  

Another class of solvable large $N$ fermionic models are those with degrees of freedom transforming as tensors under continuous symmetry groups \cite{Witten:2016iux,Klebanov:2016xxf} (for reviews, see \cite{Klebanov:2018fzb,Gurau:2019qag}). A simple example \cite{Klebanov:2016xxf} is the $O(N)^3$ symmetric quantum mechanics for $N^3$ Majorana fermions $\psi^{abc}$. In these tensor models the interaction is disorder-free, so the standard rules of quantum mechanics apply. Interestingly, the large $N$ limit is similar to that in the SYK model because in both classes of models the perturbative expansion is dominated by the ``melonic" Feynman diagrams, which can be summed \cite{Gurau:2009tw,Gurau:2011aq,Gurau:2011xq,Bonzom:2011zz,Tanasa:2011ur,Bonzom:2012hw,Carrozza:2015adg,Tanasa:2015uhr,Yoon:2017,Ferrari:2017jgw,Gubser:2018yec,Giombi:2018qgp,Klebanov:2019jup,Prakash:2019zia,Popov:2019nja}.  
Since the Hubbard and t-J models do not have any random couplings, the disorder-free tensor models may be viewed as 
their generalization,
and it is interesting to investigate if they can incorporate some interesting physical effects in a solvable setting.
One possibility is to interpret the three indices of the tensor $\psi^{abc}$, where $a,b,c=1, \ldots, N$, as labeling the sites of a 3-dimensional cubic lattice \cite{PhysRevB.100.075101}. Then the tensor models may perhaps be interpreted as non-local versions of the Hubbard model. 

It is also natural to generalize the Majorana tensor model of \cite{Klebanov:2016xxf} to the cases where the indices have different ranges: $a=1, \ldots N_1$,  $b=1, \ldots N_2$, $c=1, \ldots N_3$; then the model has 
$O(N_1)\times O(N_2)\times O(N_3)$ symmetry \cite{Bulycheva:2017ilt,Klebanov:2018nfp} (see also \cite{Ferrari:2017ryl,Ferrari:2017jgw}). 
The traceless Hamiltonian of this model is  \cite{Klebanov:2016xxf,Klebanov:2018nfp}
\begin{gather}
H = g\psi_{abc}\psi_{ab'c'}\psi_{a'bc'}\psi_{a'b'c} - \frac{g}{4} N_1 N_2 N_3\left(N_1-N_2+N_3\right)\ , \label{GenHam}
\end{gather}
where $\left\{\psi_{abc},\psi_{a'b'c'}\right\} = \delta_{aa'}\delta_{bb'}\delta_{cc'}$.
If the ranks $N_i$ are sent to infinity with fixed ratios, then the perturbation theory is dominated by the melonic graphs. However, it is also interesting to consider the cases where one or two of the $N_i$ are not sent to infinity. Such models with $O(N)\times O(2)^2$ and $O(N)^2\times O(2)$ symmetry were studied in \cite{Klebanov:2018nfp}
and were shown to be exactly solvable, with the integer energy spectrum in units of $g$. The $O(N)\times O(2)^2$ model has the familiar vector large $N$ limit, where $gN=\lambda$ is held fixed. A closely related vector model, which we also study in this paper, has Majorana variables $\psi_{aI}$, $I=1,\ldots, 4$, and symmetry enhanced to $O(N)\times SO(4)$:
\begin{equation}
H_{O(N)\times SO(4)}= \frac {g}{2} \epsilon_{IJKL} \psi_{aI}\psi_{aJ} \psi_{a'K} \psi_{a'L}\ .
\label{N4Hamilt}
\end{equation}
The 
$O(N)^2\times O(2)$ model, which may be viewed as a complex fermionic matrix model \cite{Klebanov:2018nfp}, has the `t Hooft large $N$ limit where all the planar diagrams contribute
(similar fermionic matrix models were studied in \cite{Anninos:2016klf,csordas2007cluster}).

In this paper we will carry out further analysis of the fermionic vector and matrix models. In particular, we study the large $N$ densities of states $\rho$ and 
analyze the resulting temperature dependence of the specific heat.
In the matrix model case, the density of states is smooth and nearly Gaussian, which is a rather familiar behavior. 
In the large $N$ vector models, we instead find a surprise: for a wide range of energies we find $\log \rho\approx -|E|/\lambda$ plus slowly varying terms. 
The approximately exponential growth of the density of states, discussed long ago in the context of hadronic physics and string theory \cite{Hagedorn:1965st,Atick:1988si}, leads to interesting behavior as the temperature approaches the Hagedorn temperature, $T_H=\lambda$.
In the Majorana vector models we indeed find critical behavior as the temperature is tuned to $\lambda$, with a sharp peak in the specific heat. In the formal large $N$ limit, the specific heat blows up as $(T_H- T)^{-2}$. This means that $T_H$ is the limiting temperature, and it is impossible to heat the system above it. However, at any finite $N$, no matter how large, the specific heat does not blow up, so it is possible to reach arbitrarily large temperatures. Thus, our model provides a demonstration of how the finite $N$ effects can smooth the Hagedorn transition.

In section \ref{Vecmod}, we study the $O(N)\times O(2)^2$ symmetric vector model. We find that the density of states exhibits exponential growth in a large range of energies, and match this with analytical results. In section \ref{Vecmodsym} we study a related vector model, where the symmetry is enhanced to $O(N)\times SO(4)$. In this case, we obtain simple closed-form expressions for the large $N$ density of states, free energy, and specific heat. 
In section \ref{Matmod}, we consider the fermionic matrix model with $O(N)^2\times O(2)$ symmetry and find that the spectrum now exhibits a nearly Gaussian distribution for sufficiently large $N$. In appendix A we study the structure of the Hilbert space of the above models, and derive the Cauchy identities from simple physical arguments.

\section{The $O(N)\times O(2)^2$ model} 
\label{Vecmod}

Let us consider the Hamiltonian (\ref{GenHam}) in the case $N_1=N$, $N_2=N_3=2$, so that it has $O(N)\times O(2)\times O(2)$ symmetry.  
We may think of one of the $O(2)$ symmetries as corresponding to charge, and the other $O(2)$ as the third component of spin $S_z$. The first index of $\psi^{abc}$, which takes $N$ values, can perhaps be interpreted as a generalized orbital quantum number.\footnote{We are grateful to Philipp Werner for this suggestion.} 
It will be convenient to think of the last two indices as one composite index taking four values ($I\in \left\{(11),(12),(21),(22)\right\}$). Thus, we have Majorana fermions $\psi_{aI}$ with anticommutation relations $\left\{\psi_{aI},\psi_{bJ}\right\}=\delta_{ab}\delta_{IJ}$. Hence, the Hilbert space of this problem, according to the results of the appendix, has a simple decomposition in the irreducible representations of the $SO(N)\times SO(4)$ group 
\begin{gather} \label{Hilvecstr}
\mathcal{H} = \sum_{\mu \subset 
	\mu_{\rm max}=((2)^{N/2})} [\mu]_{O(N)} \otimes [\left(\mu_{\rm max}/\mu\right)^T]_{O(4)},
\end{gather} 
where $[\mu]_{G}$ stands for a representation of the group $G$ described by the Young Tableaux $\mu$. In the 
Hilbert space of our model, the Young Tableaux of $SO(N)$ contains at most 2 columns and $N/2$ rows.
In terms of fermions $\psi_{aI}$, the Hamiltonian \eqref{GenHam} may be rewritten as
\begin{gather}
H 
= \frac{g}{2} \epsilon_{IJKL} \psi_{aI}\psi_{aJ} \psi_{a'K} \psi_{a'L} - 2 g\left [\left(\psi_{ab1}\psi_{ab2}\right)^2-\left(\psi_{a1c}\psi_{a2c}\right)^2\right ]\ .
\label{Vecsimpler}
\end{gather}
The last two terms are the charges of the two $O(2)$ groups, which break the $SO(4)$ symmetry of the first term containing the invariant tensor $\epsilon_{IJKL}$. Each of the terms has a simple action on each of the terms of \eqref{Hilvecstr},
since $O(2)\times O(2) \subset O(4)$ could be thought of as the Cartan subalgebra of $O(4)$, and we know how the Cartan subalgebra acts in the representations of $O(4)$.
The normalized generators of the $SO(4)$ group have the form
\begin{gather}
J_{IJ} =  \psi_{aI}\psi_{aJ},
\end{gather}
and can be used to split the lie algebra $\mathfrak{so}(4)$ into the direct sum of the two $\mathfrak{su}(2)$ algebras, which we have labeled by $+$ and $-$, as follows:
\begin{gather} \label{Kintro}
K^\pm_1 = \frac12 J_{01}\pm \frac{1}{2} J_{23},\quad 
K^\pm_2 = \frac12 J_{02}\pm \frac{1}{2} J_{31},\quad
K^\pm_3 = \frac12 J_{03}\pm \frac{1}{2} J_{12}.
\end{gather}
It is easy to see that both sets $K^+_i$ and $K^-_i$ comprise an $SU(2)$ algebra,
and thus the representations of the two $SU(2)$ groups with spins $Q_+/2$ and $Q_-/2$, respectively, fully determine the representation of the $SO(4)$ group.
One can derive the following algebraic relation: 
\begin{gather}
\frac {g}{2} \epsilon_{IJKL} \psi_{aI}\psi_{aJ} \psi_{a'K} \psi_{a'L} = \frac{g}{2}\epsilon_{IJKL} J_{IJ} J_{KL} = \notag\\
= 4 g \sum_i \left[ \left(K^+_i\right)^2 - \left(K^-_i\right)^2\right] = g\left[Q_+(Q_+ + 2) - Q_- (Q_- + 2) \right],
\end{gather}
where we have used that $ \left(K^+_i\right)^2$ is the quadratic Casimir operator and we know its value in each of the representations of $SU(2)$.
It is also interesting to notice that from \eqref{Kintro} we have 
\begin{gather}
\psi_{ab1}\psi_{ab2} = 2 K^+_1  ,\quad \psi_{a1c}\psi_{a2c} = 2  K^-_1.
\end{gather}
This allows one to rewrite the Hamiltonian only in terms of the $SO(4)$ representations. If we have a representation with $SU(2)$ spins $(Q_+/2,Q_-/2)$, then all eigenvectors with definite $K^\pm_1$ are the eigenvalues of Hamiltonian with energies
\begin{gather}
E(Q_+,Q_-,q_+,q_-) =g \left[Q_+(Q_+ + 2) - Q_- (Q_- + 2) + 2 q_{-}^2 - 2 q_{+}^2 \right],\notag\\
 K^\pm_1 \ket{Q_\pm, q_\pm} = q_\pm \ket{Q_\pm,q_\pm}\ .\label{exactsolvecmod}
\end{gather}
The degeneracy of such a state is determined by the dimension of the corresponding $SO(N)$ representation. Because we know the structure of the Hilbert space \eqref{Hilvecstr}, we can determine the complete structure of the spectrum. If we have a $SO(N)$ representation with a Young tableaux $\mu$ consisting of two columns of the length $\mu_{1}\geq \mu_2 \geq 0$, the corresponding representations of $SO(4)$ have $Q_+ = N-\mu_1-\mu_2, Q_-= \mu_1 - \mu_2$, and the dimension of the representation of $SO(N)$ is \cite{cvitanovic2008group} 
\begin{gather}\label{dimenofvecmod}
\dim\left(Q_+, Q_-\right) = \frac{(Q_+ +1) (Q_- + 1) N! (N+2)!}{\left(\frac{N- Q_+ - Q_-}{2}\right)! \left(\frac{N+ Q_+ - Q_-+2}{2}\right)! \left(\frac{N- Q_+ + Q_- + 2}{2}\right)! \left(\frac{N + Q_+ + Q_- + 4}{2}\right)!}.
\end{gather}
From this one can see that each set of pairs of non-negative integers $(Q_+, Q_-)$ whose sum is constrained to take values $N, N-2, N-4, \ldots$ appears once.
This formula allows us to study the density of states in the vicinity of the ground state and of $E=0$. 

\begin{figure}
	\centering
	\includegraphics[scale=1.25]{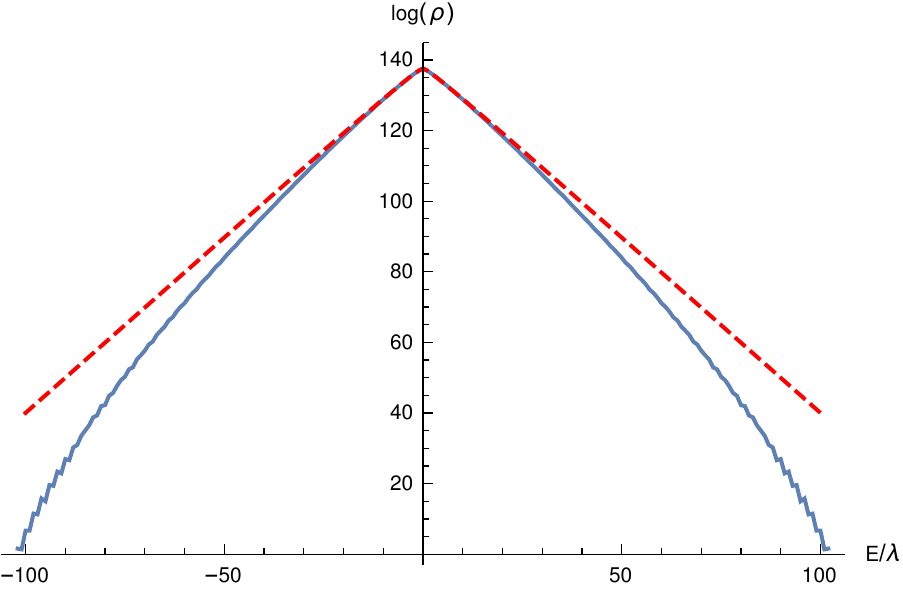}%
	\caption{The logarithm of the density of states of the $O(N)\times O(2)^2$ vector model, shown for $N=100$. 
For comparison, the large $N$ result (\ref{vecdenan}) is shown with a dashed line.}
	\label{cuspfig}
\end{figure}

The ground state ($ E_0 = - g N(N+2) $) corresponds to the choice of $Q_+ = 0, Q_-=N$, thus $q_+ \equiv 0 $ and the spectrum in its vicinity has the form,
\begin{gather} \label{simsec}
E = 2 g q_-^2  - g N(N+2),\quad \deg =\dim(N,0)= 1,\quad  -N \leq q_- \leq N.
\end{gather}
The states immediately above the ground state are labeled by $q_-$ and the gap between them is of the order $g\sim\frac{\lambda}{N}$. The next excited states correspond to the choice $Q_+ >0$. The gap between such states and the ground state is of the order $\Delta E \sim g N \sim \lambda$ and is finite in the large $N$ limit, but the dimension of the representation is of the order $\dim \sim N^{Q_+}$ and diverges in the large $N$ limit. Immediately above the ground state ($\delta E \sim \lambda$, $Q_+ = 0$) the density of states may be approximated as 
\begin{gather}
\Gamma(E)=\left\{\text{\# of states: } E_{\rm st}\leq E+E_0\right\} =\left\{\text{\# of $q_-$}:2 g q_-^2 - g N(N+2)\leq E+E_0\right\}\approx \sqrt{\frac{E}{2 g}},\notag\\
\rho(E) = \frac{d\Gamma}{dE} \sim \sqrt{\frac{1}{8 g E}}\ ,\quad E \sim \frac{\lambda}{N}\ .
\end{gather}
On the other hand, near $E=0$, the logarithm of the density of states exhibits an unusual cusp-like behavior shown in figure \ref{cuspfig}. Another remarkable feature is its approximately linear behavior for a large range of energies.

For $|E|/\lambda$ of order $1$, the dominant contributions come from the states with large charges $Q_\pm \sim \sqrt{N}\gg 1$. 
In this regime we can apply the Stirling approximation to the factorials in (\ref{dimenofvecmod}) to obtain 
\begin{gather}
\dim(Q_+,Q_-) \approx  2^{2N}Q_+ Q_- \exp \left (- \frac{Q_+^2 + Q_-^2}{N}\right )\ .
\label{logdim}
\end{gather}
To obtain the density of states in the large $N$ limit,  
we introduce the variables $t_\pm = \frac{Q_\pm}{\sqrt{N}}\ , u_\pm= \frac{q_\pm}{\sqrt{N}}$, and $x=\frac{E}{\lambda}$. Then we have  
\begin{gather}\label{complexintegralvecmod}
\rho(x) \sim \int\limits^\infty_0 t_+ dt_+ \int \limits^\infty_0 t_- dt_- e^{-t_-^2 - t_+^2 } \int\limits^{t_+}_{-t_+} du_+ \int \limits^{t_-}_{-t_-} du_- \delta\left(x + t_+^2 - t_-^2 + 2 u_-^2 - 2 u_+^2\right)\ .
\end{gather}
This may be evaluated if we  first perform the integrals over $T_\pm = t^2_\pm$:
\begin{gather}
\rho(x) \sim \int\limits^\infty_{-\infty} du_+ \int\limits^\infty_{-\infty} du_-  \int \limits^\infty_{u_+^2} d T_+
\int \limits^\infty_{u_-^2} d T_-
 e^{-T_- - T_+} \delta\left(x + T_+ - T_- + 2u_-^2 - 2u_+^2\right)  \sim\notag\\ \sim
\int\limits^\infty_0\, du\, e^{-2 u^2-|x|}\,\sqrt{|x|+u^2}+\int\limits^\infty_{\sqrt{|x|}}\, du\, e^{|x|-2 u^2}\,\sqrt{u^2-|x|}= \notag\\
= e^{-|x|} {}_1F_1\left(-\frac12;0;2 |x|\right)+\frac{e^{|x|}}{\sqrt{2}} \sqrt{|x| }G^{0,1}_{1,2}\left( {}_{-\frac12, \frac12}^{\phantom{\frac12}1} \bigg| 2|x|  \right) \label{vecdenan}
\ , \end{gather}
where the last term involves the Meijer $G$-function.
The formula \eqref{vecdenan} is in good agreement with the numerical results (see figure \ref{cuspfig}). Expanding $\rho(x)$ near $x=0$ we see that %{\bf The factor $e^{-|x|}$ below is surprising}
\begin{gather}\label{rhozerovecmod}
\rho(x) \sim 1 +  \frac14 \left(2 \log \frac{|x|}{2} + 2 \gamma  - 1\right)x^2\ ,
\end{gather}
which exhibits a singularity at $x=0$:
$\rho''(0)$ diverges, signaling a breakdown of the Gaussian approximation of the density of states.
 We also note that, for $x \gg 1$, $\rho(x) \sim |x|^\frac12 e^{-|x|}$.

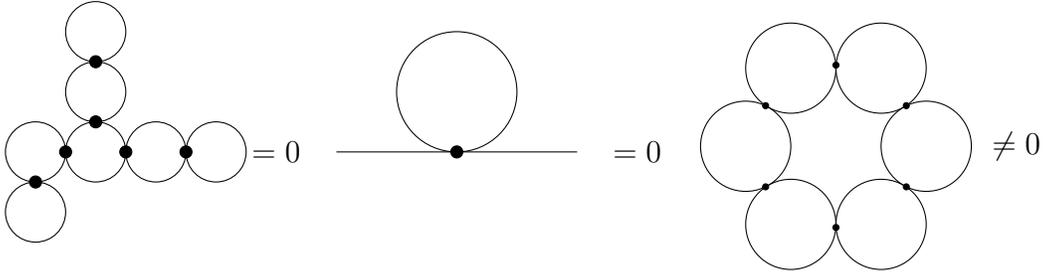
\begin{figure}
	\centering
	\begin{tikzpicture}[scale=0.8]
	\draw[fill] (0,0) circle (1mm);
	\draw (0,0.5) circle (0.5);
	\draw (0,-0.5) circle (0.5);
	\draw[fill] (0,-1) circle (1mm);
	\draw (0,-1.5) circle (0.5);
	\draw[fill] (-0.5,-1.5) circle (1mm);
	\draw[fill] (0.5,-1.5) circle (1mm);
	\draw (1,-1.5) circle (0.5);	
	\draw[fill] (1.5,-1.5) circle (1mm);
	\draw (2,-1.5) circle (0.5);
	\draw (-1,-1.5) circle (0.5);
	\draw[fill] (-1,-2) circle (1mm);
	\draw (-1,-2.5) circle (0.5);
	\node at (3,-1.5) {$=0$};
	
	\draw (4,-1.5)--(6,-1.5);
	\draw[fill] (6,-1.5) circle (1mm);
	\draw (6,-0.5) circle (1);
	\draw (8,-1.5)--(6,-1.5);
	
	\node at (9,-1.5) {$=0$};
	
	\begin{scope}[xshift=350,scale=0.5,yshift=-80]
	\def \n {6}
	\def \radius {3}
	
	\foreach \s in {1,...,\n}
	{
		\draw ({360/\n * (\s - 1)}:\radius) circle (1.5);
		\draw[fill] ({360/\n * (\s - 1)+30}:2.7) circle (1mm);
	}
	\node at (6,0) {$\neq 0$};
	\end{scope}
	\end{tikzpicture}
	\caption{The cactus diagrams, which are of order $N$, vanish due to the Majorana nature of the variables. The ``necklace" diagrams, are not equal to zero and give the leading 
contributions in the large $N$ limit, which are of order $N^0$.}
	
	\label{fig:treevector}
\end{figure}
We can present an argument for why the density of states is not Gaussian near the origin. The high temperature expansion of the free energy is:
\begin{gather}\label{polyvector}
\tr e^{-\beta H} = e^{-F},\quad F = \sum\limits^\infty_{n=1} (-1)^{n+1} \beta^n \tr_{\rm con}  \left[H^n\right]\ .
\end{gather}
The quantity on the right-hand side of \eqref{polyvector} may be computed with the use of Feynman diagrams.
For vector models, the ``cactus" or ``snail" diagrams, shown in figure \ref{fig:treevector}, typically dominate in the large $N$ limit \cite{Klebanov:2018fzb,MaldacenaTasi}. 
However, in our problem they vanish due to the Majorana nature of the variables. Therefore, for any connected part, the trace begins with the subleading term
\begin{gather}
\frac{1}{N^n}\tr_{\rm con} \left[ H^n \right] = N^0 C_1 + N^{-1} C_2+\ldots 
\end{gather}
It is easy to see that $C_1$ comes from the necklace diagrams in figure \ref{fig:treevector}, which give
\begin{gather}
C_1 = \sum_{k=1}^\infty \frac{(g N)^k}{k} \frac{(1+(-1)^k)}{2},
\end{gather}
where the factor of $\frac{1}{k}$ comes from the symmetries of the necklace diagrams. These necklace diagrams may be interpreted as trajectories of a particle propagating in one dimension. Introducing the `t-Hooft coupling $\lambda=g N$ and taking the large $N$ limit while keeping $\lambda$ finite, we calculate the free energy,
\begin{gather} \label{FreeVectorEnergy}
F = \sum^\infty_{k=1} \frac{(\beta \lambda)^k}{k}\frac{(1+(-1)^k)}{2} = -\frac{1}{2} \left( \log(1+\beta \lambda) + \log(1-\beta \lambda) \right)= -\frac{1}{2}\log\left[1-(\beta \lambda)^2\right].
\end{gather}
The inverse Laplace transformation with respect to $\beta$ yields the density of states $\log \rho(E) \sim a - \frac{|E|}{\lambda}$.
From this one can derive that the distribution must have a Laplace-like form, and this agrees with the numerical results.

\begin{figure}
	\centering
\includegraphics[width=0.9\textwidth]{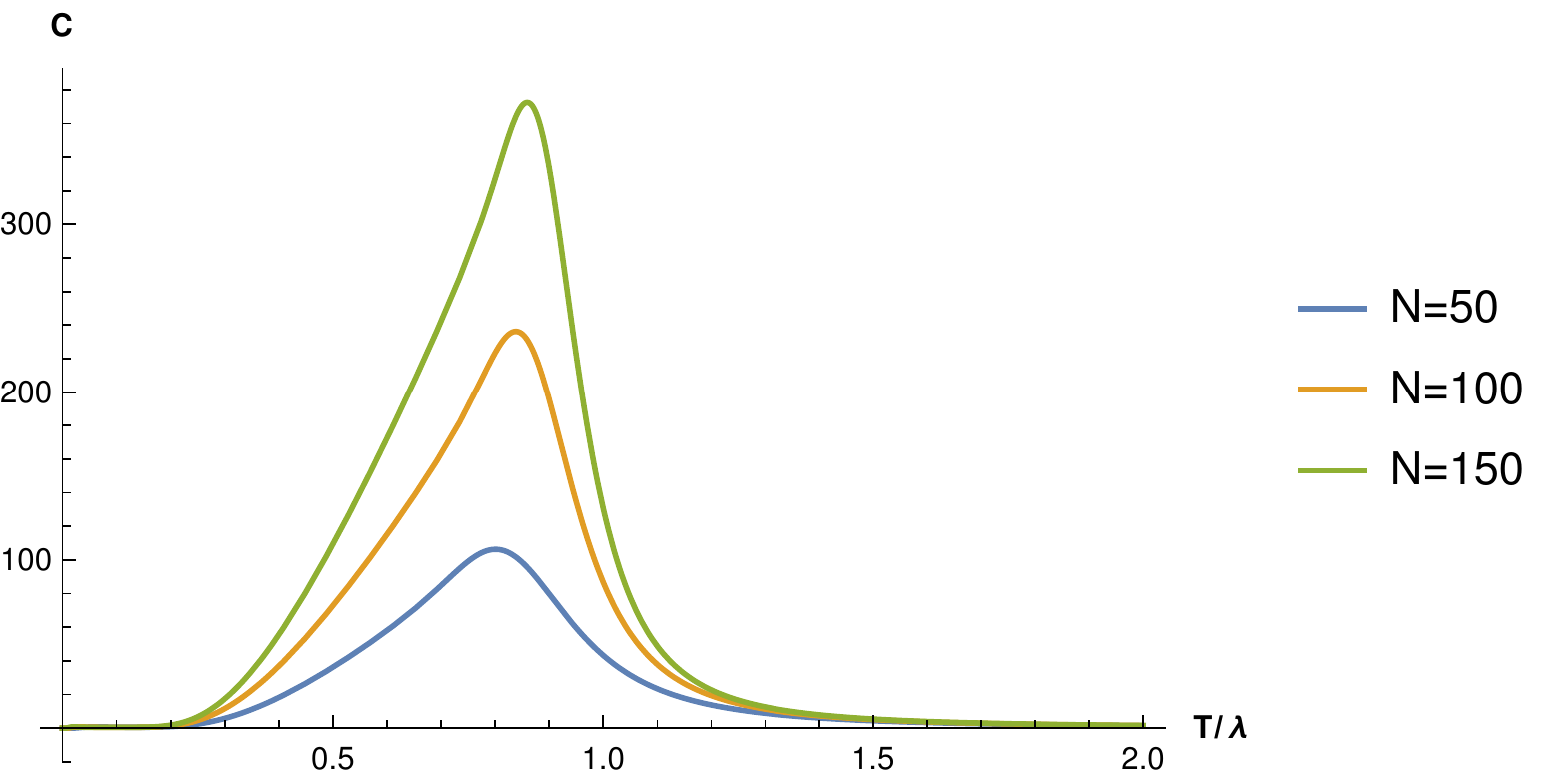}
	\caption{The plot of specific heat $C$ for the $O(N)\times O(2)^2$ model, as a function of temperature $T/\lambda$, for $N=50,100,150$. 
		The specific heat has a pronounced peak which gets closer to $T/\lambda=1$ as $N$ grows. 
	}
	\label{fig:SpecHeat}
\end{figure}

Let us review the physical effects of the approximately exponential behavior of $\rho$. In the canonical ensemble, the partition function as a function of inverse temperature $\beta$ is
\begin{gather}
Z= \int_0^\infty d\tilde E \rho(\tilde E) e^{-\beta \tilde E} \ ,
\end{gather}
where we define $\tilde E= E-E_0$ to be the energy above the ground state. If $\rho(\tilde E)\sim e^{\tilde E/T_H}$, then $Z$ diverges for $\beta< \beta_H$, where $\beta_H=1/T_H$; this is the well-known Hagedorn behavior. 
For our vector model, the Hagedorn temperature is $T_H=\lambda$. However, the divergence is cut off by the fact that $\rho (\tilde E)$ grows approimately exponentially only from some initial value $\tilde{E}_0$ up to some critical value $\tilde E_c$, as shown in figure \ref{cuspfig}.
The contribution to $Z$ from this region of energies is
\begin{gather}
Z_{\rm Hagedorn}\sim {e^{- (\beta- \beta_H) \tilde{E}_0}- e^{- (\beta- \beta_H) \tilde E_c}\over \beta - \beta_H}.
\end{gather} 
The presence of the denominator produces a logarithmic term in the free energy, but it is cut off by the numerator before it diverges. It follows that
the specific heat $C=-T\partial^2 F/\partial T^2$ may be approximated
by
\begin{gather}
C=\frac{1}{\left(\frac{T}{T_H} - 1\right)^2 } + \frac{\delta \tilde{E}^2}{4 T^2 \sinh^2\left(\frac{\delta \tilde E}{2}\left[\frac{1}{T} - \frac{1}{T_H}\right]\right)}, \quad \delta \tilde E = \tilde{E}_c -\tilde{E}_0, \label{specheatvec}
\end{gather}
where $\delta \tilde E $ goes to infinity in the large $N$ limit and the second term vanishes. 
Thus, for large enough $N$, there should be a clear peak in the specific heat. 
This simple analytic argument for the existence of a peak is supported by the numerical plots of specific heat shown in figure \ref{fig:SpecHeat}.
For any finite $N$, the height of the peak in $C$ is finite, so that it is possible to heat the system up to any temperature. However, in the formal large $N$ limit, the specific heat blows up as
$(T-T_H)^{-2}$ so the Hagedorn temperature is the limiting temperature. This shows that the finite $N$ effects smooth out the Hagedorn transition.

\section{The $O(N)\times SO(4)$ model}
\label{Vecmodsym}

In this section we study the simpler vector model where we retain only the first term in the Hamiltonian \eqref{Vecsimpler}. The symmetry is then enhanced to $O(N)\times SO(4)$ symmetry. 
 Since $SO(4)\sim SU(2)\times SU(2)$, we can think of one of the $SU(2)$ groups as corresponding to the spin of the fermions.  
From the previous section we know that the spectrum of the model may be expressed in terms of the two $SU(2)$ spins, $Q_\pm/2$, where $Q_\pm$ are non-negative integers 
whose sum is constrained to take values $N, N-2, N-4, \ldots$. The energies and their degeneracies are:
\begin{gather}
E(Q_+,Q_-) = g\left[Q_+(Q_++2) - Q_-(Q_-+2)\right] = g (Q_+ - Q_-) (Q_+ + Q_- + 2)\ , \notag\\
\deg(Q_+ ,Q_-) = \frac{(Q_+ +1)^2 (Q_- + 1)^2 N! (N+2)!}{\left(\frac{N- Q_+ - Q_-}{2}\right)! \left(\frac{N+ Q_+ - Q_-+2}{2}\right)! \left(\frac{N- Q_+ + Q_- + 2}{2}\right)! \left(\frac{N + Q_+ + Q_- + 4}{2}\right)!}\ .
\label{newfact}
\end{gather}
The ground state corresponds to $Q_+ = 0, Q_-=N$; it has energy $E_0=-\lambda (N+2)$ and degeneracy $N+1$. 
For the series of states $Q_+= m$, $Q_-=N-m$, where $m$ are positive integers much smaller than $N$, we find the excitation energies $E_m- E_0\approx 2m \lambda$. These states are equally spaced in the large $N$ limit, and their degeneracies behave for large $N$ as 
$\frac {N^{1+m}}{(m+1)!}$. Thus, the density of states $\rho(E)$ near the lower edge grows as $\sim N^{\frac{E-E_0}{2\lambda}}$. This edge behavior does not have a smooth large $N$ limit; it is very different from the random matrix behavior $\sim \sqrt{E-E_0}$ which is observed in the SYK model.

\begin{figure}
	\centering
	\includegraphics[width=8cm]{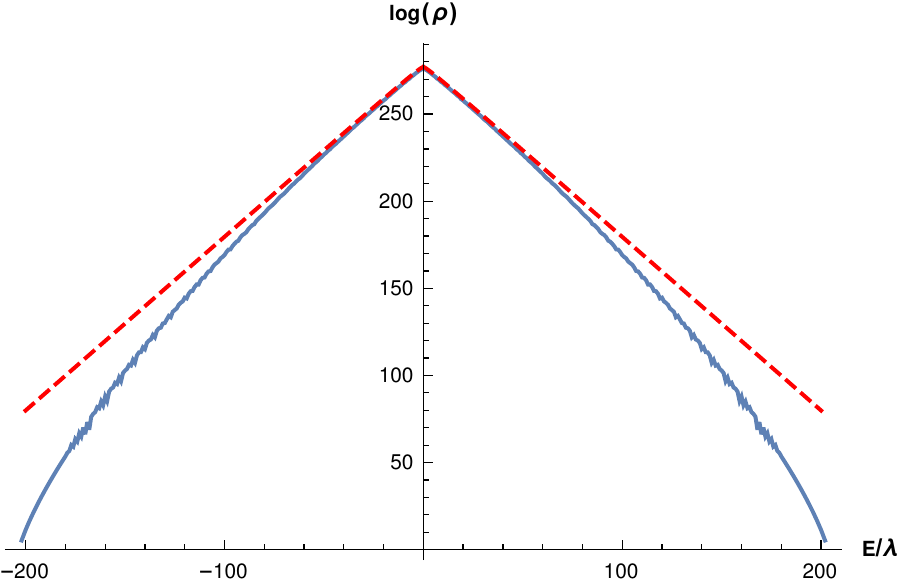} % 
	\hfill
	\includegraphics[width=8cm]{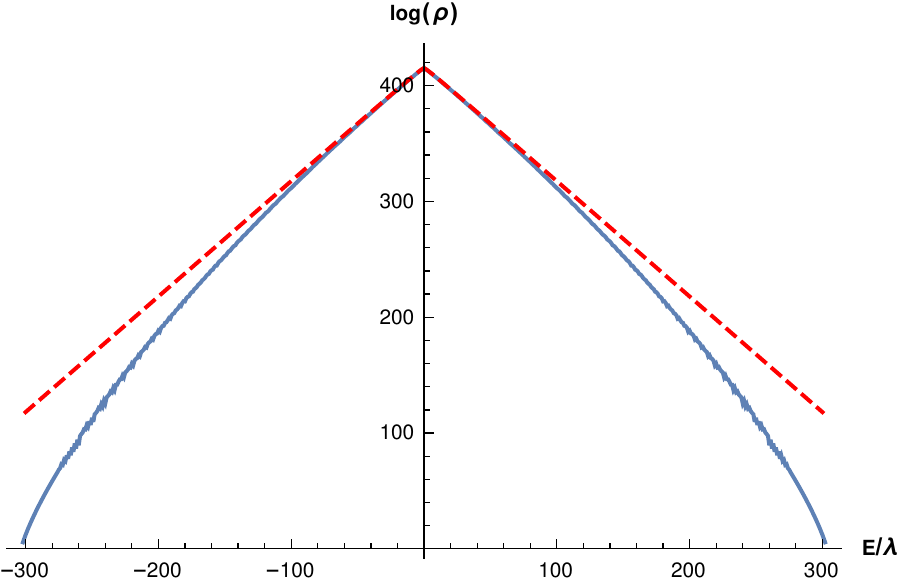}
	\caption{The logarithm of the density of states for the $O(200)\times SO(4)$ (on the left) and $O(300) \times SO(4)$ (on the right) models with Hamiltonian (\ref{N4Hamilt}). 
For comparison, the large $N$ result (\ref{Bess}) is shown with a dashed line.
}
\label{simpvecdens}
\end{figure}

Just like for the $O(N)\times O(2)^2$ model, we find that the large $N$ limit of the $O(N)\times SO(4)$ model has a nearly linear behavior of the logarithm of density of states for a certain range of $E/\lambda$ (see figure \ref{simpvecdens}).   
Let us study this function more precisely near the middle of the distribution, following the procedure used in the previous section.  We include the contributions of representations where $Q_\pm\sim \sqrt N$, and 
introduce variables $x_\pm = Q_\pm /\sqrt N$.
The energy is then given by $E = \lambda \left(x_+^2 - x_-^2\right)$. Using the Stirling approximation for the factorials in (\ref{newfact}), we find that 
the density of states  is
\begin{gather}
\rho(E) \sim \int_0^\infty dx_+  \int_0^\infty dx_- x_+^2 x_-^2 e^{-(x_+^2+ x_-^2)}\delta \left(E - \lambda \left(x_+^2-x_-^2\right)\right) \ .
\end{gather}
This integral can be evaluated in closed form:
\begin{gather}
\rho(E) = 2^{2N} 
\frac{|E|}{\pi \lambda^2} K_1 \left (\frac{|E|}{\lambda}\right ) \ , 
\label{Bess}
\end{gather}
where $K_1$ is the modified Bessel function, and the normalization is such that $\rho$ integrates to the total number of states, $2^{2N}$. 
Plotting (\ref{Bess}), we see that in the range where $N^{-1}\ll |E|/\lambda\ll N$, it is close to the numerical results in figure \ref{simpvecdens}. 
The expansion of (\ref{Bess}) near the origin,
\begin{gather}
\rho = 2^{2N} \frac{1}{\pi \lambda}\bigg (1+ \frac14 (2\log \frac{|x|}{2} + 2\gamma-1) x^2 + O(\log |x| x^4) \bigg )\ , \quad x=\frac{E}{\lambda}\ ,
\end{gather}
shows that $\rho''(0)$ diverges.
The reasons for this unusual behavior in the large $N$ limit were discussed in the previous section. We also note that $\rho \sim |x|^{1/2} e^{-|x|}$ for $|x|\gg 1$. 

\begin{figure}
	\centering
	\includegraphics[width=0.9\textwidth]{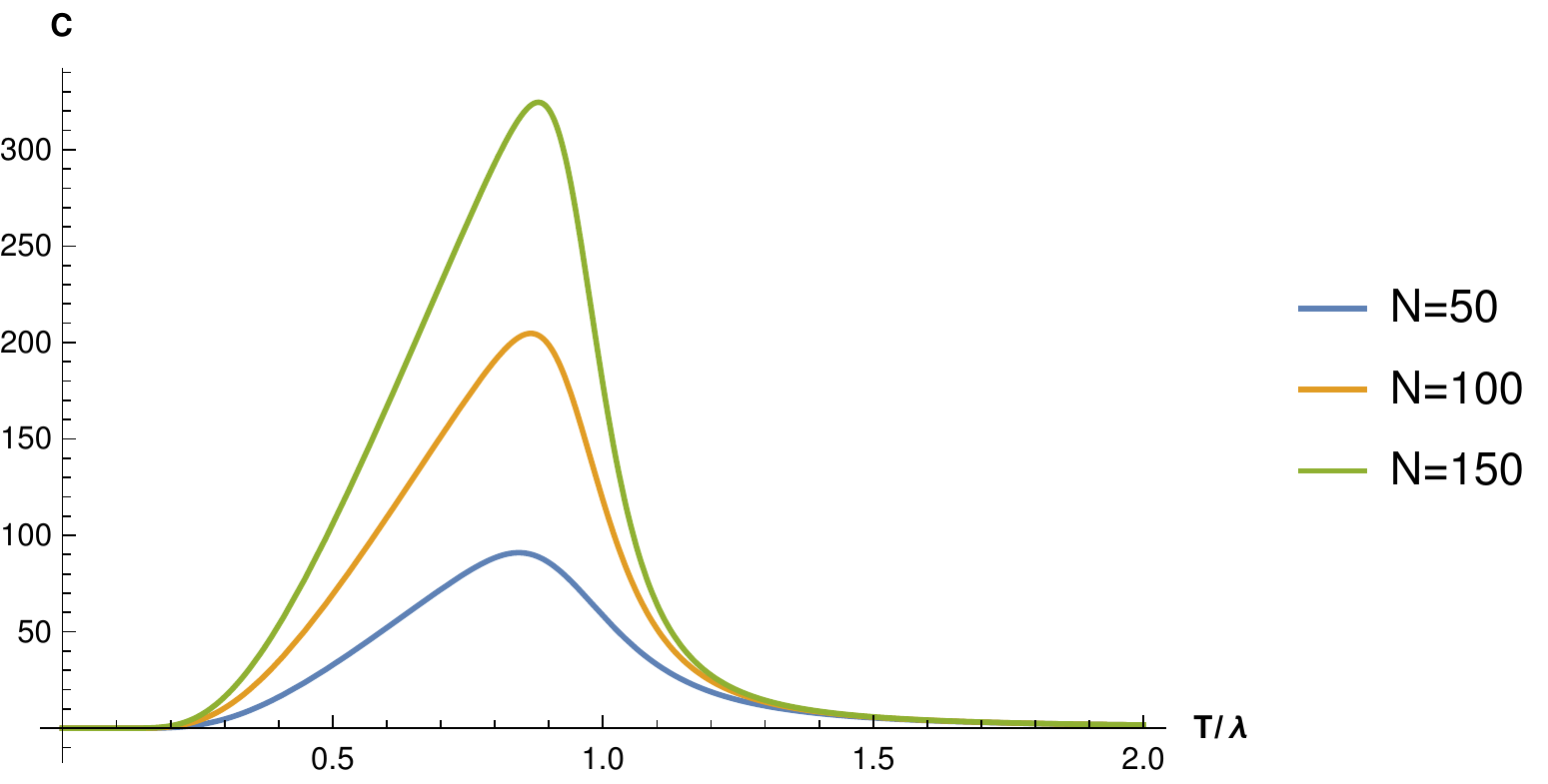}
	\caption{The plot of specific heat $C$ for the $O(N)\times SO(4)$ model, as a function of temperature $T/\lambda$, for $N=50,100,150$. 
		The peak in specific heat gets closer to $T/\lambda=1$ as $N$ increases. 
	}
	\label{fig:SpecHeatNew}
\end{figure}

The approximation \eqref{Bess} can be used to get the large $N$ limit of the free energy: 
\begin{gather}
 F(T) = -T \log Z(T) =\frac32  T \log \left( \frac{\lambda^2}{T^2} - 1\right),
\end{gather}
up to an additive term linear in $T$.
The specific heat diverges at the Hagedorn temperature $T_H=\lambda$, 
\begin{equation} 
C(T) = - T \frac{\partial^2 F}{\partial T^2}= \frac{3 \lambda^2 \left(T^2+\lambda^2\right)}{\left(T^2-\lambda^2\right)^2} 
\ .
\end{equation}
Note that this is of order $N^0$ for $T<T_H$, as usual for the Hagedorn transition.
For a finite $N$, the divergence is cut off, but the peak is prominent; see
figure \ref{fig:SpecHeatNew}.

We can write the Hamiltonian (\ref{N4Hamilt}) in terms of complex fermions by introducing the following operators:
\begin{align} 
& c_{a1} = \frac{1}{\sqrt{2}}\left(\psi_{a1}+i \psi_{a2}\right),\qquad 
\bar{c}_{a1} = \frac{1}{\sqrt{2}}\left(\psi_{a1}-i \psi_{a2}\right)\ ,\notag\\
& c_{a2}=\frac{1}{\sqrt{2}}\left(\psi_{a3}+i \psi_{a4}\right),\qquad  \bar{c}_{a2}=\frac{1}{\sqrt{2}}\left(\psi_{a3}-i \psi_{a4}\right)\ . \label{sff}
\end{align}
We may think of $a=1, \ldots N$ as a 1-dimensional lattice index, so that there are two complex fermions at each lattice site.
The lattice Hamiltonian is then non-local:\footnote{This Hamiltonian should be contrasted with the local fermionic $O(N)$ chains (see, for example, the recent paper \cite{Hakobyan:2020yoy}), where there are $N$ fermions at each lattice site.}  
\begin{gather}
H_{O(N)\times SO(4)} = -\frac{g N}{2} - \frac{g N^2}{4} + g \bar{c}_{a1}\bar{c}_{a2}c_{b1}c_{b2} + g\left(\sum_a \vec{J}_a\right)^2\ , 
\qquad \vec{J}_a = \bar{c}_{a\alpha} \vec{\sigma}_{\alpha \beta} c_{a \beta}\ . 
\label{N4Hamilta}
\end{gather}
It is then not surprising that this model exhibits a phase transition in the large $N$ limit: it corresponds to the limit where the lattice becomes infinitely long. 

%which reveals the essentially non-interacting nature of the model. It may be thought of as free hopping of a composite object $\bar{A}_a=\bar{c}_{a1}\bar{c}_{a2}$, consisting of two fermions with opposite spins, with the index $a$ considered as a spatial index.

For the Hilbert space of the model containing fermions $\psi_{i J}$, the quadratic Casimirs of the $SO(N)$ and $SO(4)$ symmetry groups satisfy the constraint \cite{Klebanov:2018nfp},
\begin{equation} 
C_2^{SO(N)} + C_2^{SO(4)}= N \left (\frac {N} {2} + 1 \right ).
\label{Casimirconstr} 
\end{equation}
In later sections we will be interested in the $SO(N)$ invariant states, and (\ref{Casimirconstr})
implies that these states must have $C_2^{SO(4)}= N\left (\frac {N} {2} + 1 \right )$. The corresponding representations of $SU(2)\times SU(2)$ have spins
$j_+=0, j_-= N/2$ or $j_+=N/2, j_-=0$. The first set of $N+1$ states has the lowest energy, while the second set of $N+1$ states has the highest energy.
In total there are $2N+2$ states which are $SO(N)$ invariant.

We may also work in terms of complex fermions $c_{ai}$,
(\ref{sff}), which are naturally acted on by $SU(N)\times SU(2)\times U(1)$. The $SU(N)$ acts on the first index, $SU(2)$ on the second, and $U(1)$ by overall phase rotation. On the Hilbert space constructed this way, the 
quadratic Casimirs satisfy the constraint \cite{Klebanov:2018nfp}
\begin{equation} 
C_2^{SU(N)} + C_2^{SU(2)}= \frac{N+2}{4N} (N^2- Q^2)\ ,
\label{Casimirconstrnew} 
\end{equation}
where $Q$ is the $U(1)$ charge.
This implies that the $SU(N)$ invariant states with $Q=0$ must be in the spin $N/2$ representation of $SU(2)$. Therefore, there are $N+1$ such states. 
There are also two $SU(N)\times SU(2)$ invariant states, which have $Q=\pm N$. Thus, the total number of $SU(N)$ invariant states is $N+3$. 

We can generalize such a model to the case of $O(N)\times SO(2M)$ with the Hamiltonian
\begin{gather}
H=i^M\frac{g}{M!}\eps_{j_1 \ldots j_{2M}} \psi_{a_1 j_1} \psi_{a_1 j_2} \ldots \psi_{a_M j_{2M-1}} \psi_{a_M j_{2M}} .
\end{gather}
This may be expressed via the higher Casimirs operators of the $SO(2M)$ group. For the case of $M=1$ we would have a simple model $O(N)\times SO(2)$,
\begin{gather}
H = i g \eps_{i j} \psi_{a i} \psi_{a j} =  2 i g  \psi_{a 1}\psi_{a 2} = 2 g \left(\bar{c}_a c_a -\frac{N}{2}\right),\quad c_a = \frac{\psi_{a1}+ i \psi_{a 2}}{\sqrt{2}}.
\end{gather}
The spectrum consists of half-integers running from $E = -\frac{N}{2}+q$  and the degeneracy $\deg(E) = {{N} \choose {q}}$ corresponds to the representation of the fully antisymmetric tensors.

\section{Fermionic matrix models}
\label{Matmod}

In this section we study the fermionic matrix models with $O(N_1)\times O(N_2)\times O(2)$ symmetry \cite{Klebanov:2018nfp}. They contain 
$2N_1 N_2$ Majorana fermions that are coupled by the Hamiltonian
\begin{gather}
H = g \psi_{abc}\psi_{ab'c'}\psi_{a'bc'}\psi_{a'b'c} - \frac{g}{2} N_1 N_2\left(N_1-N_2+2\right) \ . \label{HamAC}
\end{gather}  
The direct numerical diagonalization of this Hamiltonian is hampered by the exponential growth of the dimension of Hilbert space as $2^{N_1 N_2}$. For $N_1=N_2=6$ it is $\approx 7\cdot10^{10}$, while for $N_1=N_2=8$ it is $\approx 2\cdot10^{18}$ states. For the former we were able to carry out Lanczos diagonalization giving the wave functions and energies of the lowest few states.
 \begin{figure}[h]
 	\centering
 	\includegraphics[width=8cm]{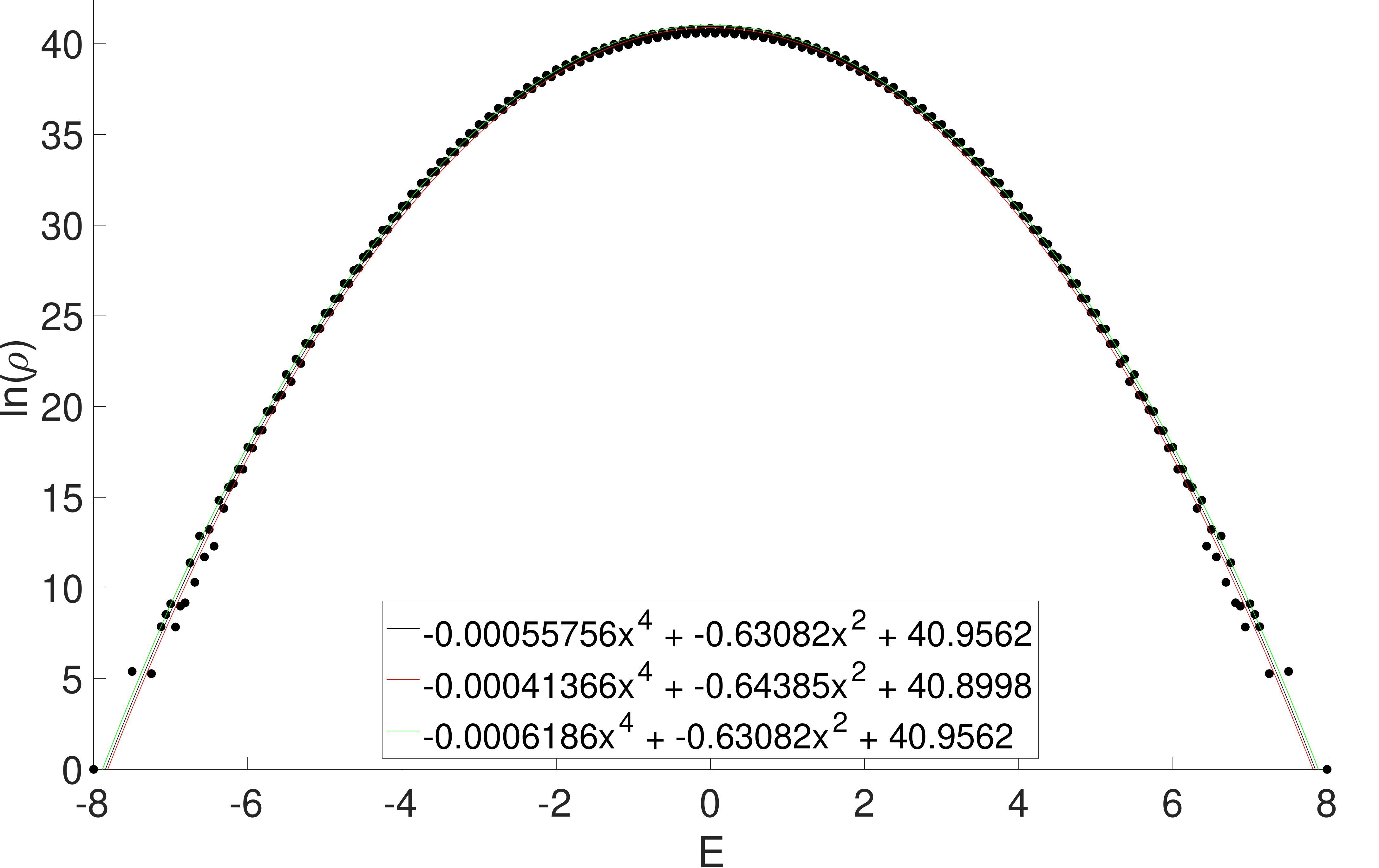}%
 	\includegraphics[width=8cm]{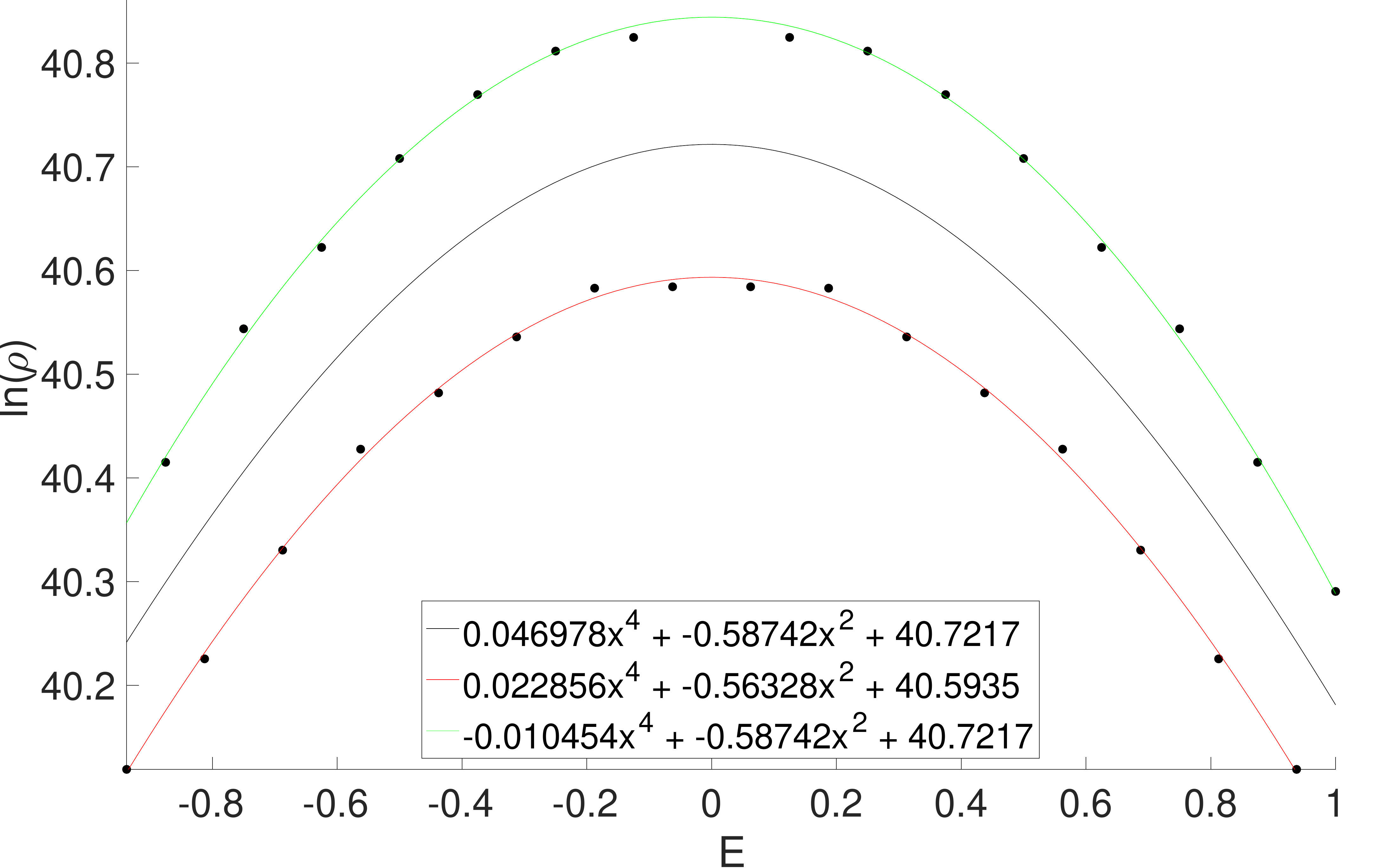}\\%
 	\includegraphics[width=8cm]{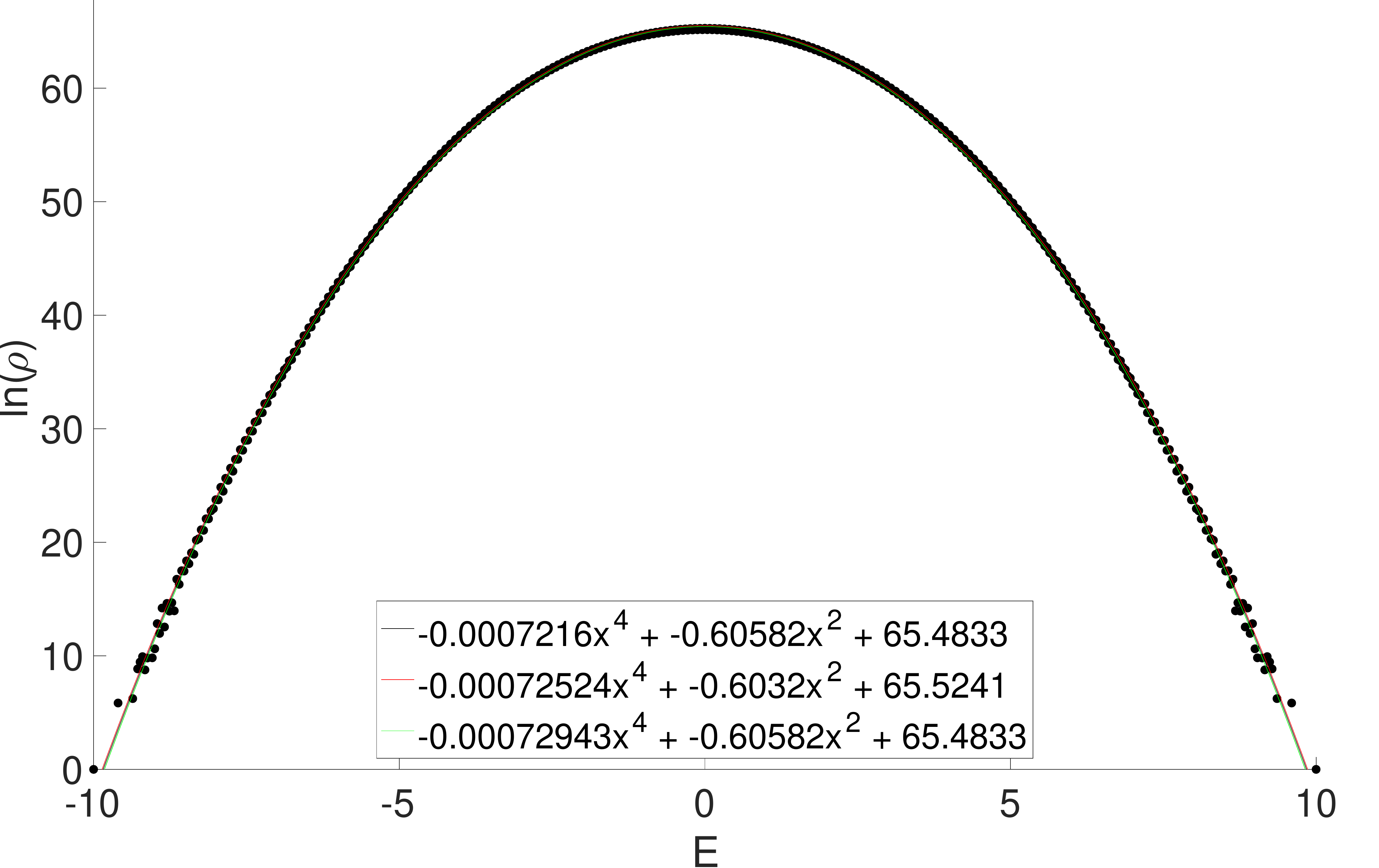}%
 	\includegraphics[width=8cm]{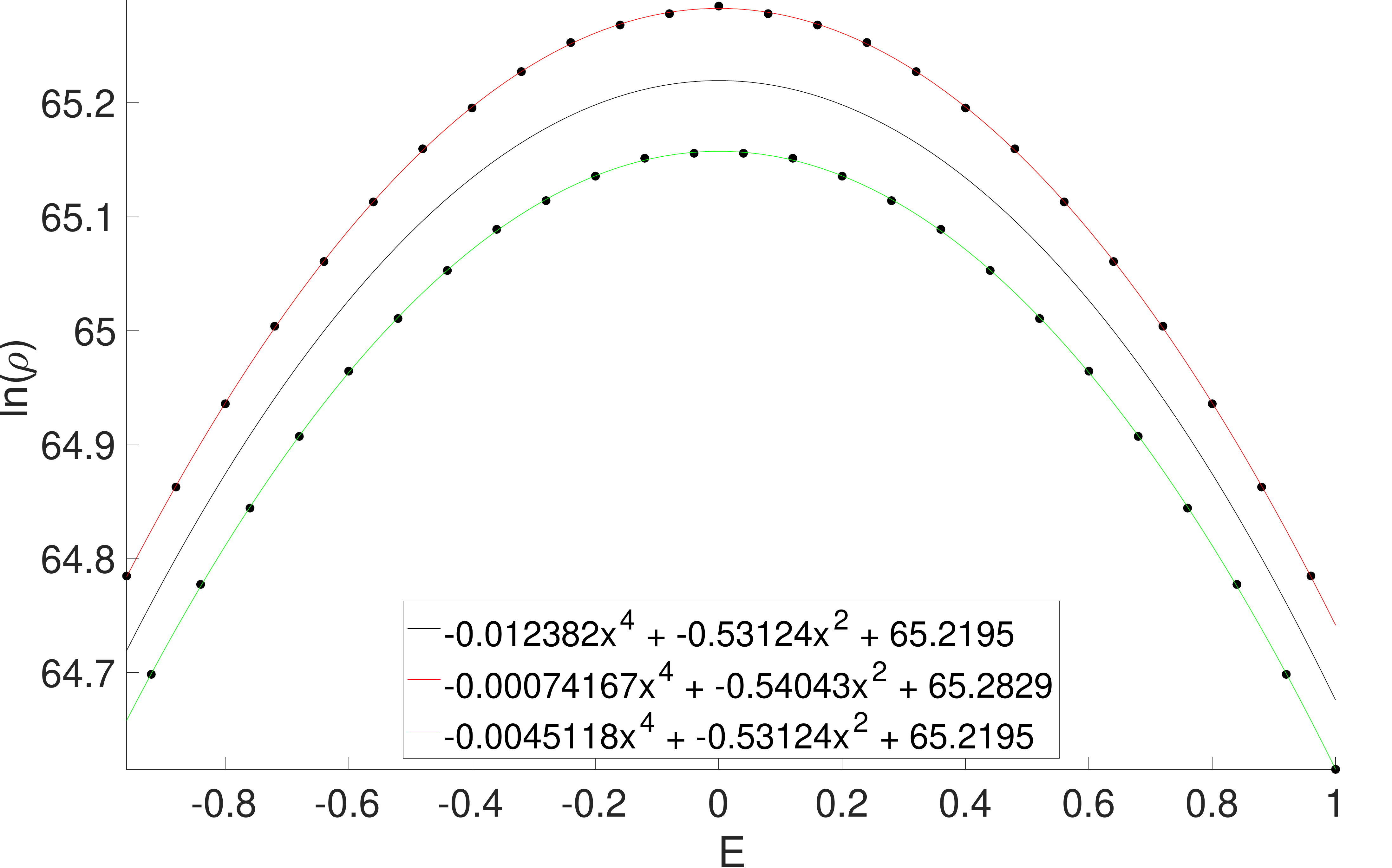}%	
 	\caption{The spectrum for $N_1=N_2=8$ and $N_1=N_2=10$ on the top and the bottom row. One can see that the spectrum is Gaussian, but split into two branches. The fit is quite close to the theoretical predictions.}
 	
 	\label{fig810}
 	
 \end{figure}
 
Fortunately,
the Hamiltonian (\ref{HamAC}) may be expressed in terms of the $U(1)$ charge $Q$, the Casimir operators of the $SO(N_i)$ symmetry groups, as well as of the $SU(N_1)$ group which acts on the spectrum \cite{Klebanov:2018nfp}:
\begin{equation} 
H= - 2 g \Bigg( 4 C_2^{SU(N_1)}  -  C_2^{SO(N_1)} +  C_2^{SO(N_2)}+ \frac{2}{N_{1}} Q^2 + (N_2- N_1) Q
-\frac{1}{4} N_{1}N_{2} (N_{1}+N_{2})   \Bigg)\ . \label{h_final}
\end{equation}
This analytical expression allows us to proceed to higher values of $N_i$. In general, all the energy eigenvalues are integers in units of $g$, but finding their degeneracies requires some calculations via the group representation theory. 

\begin{figure}
	\centering
	\includegraphics[width=8.5cm]{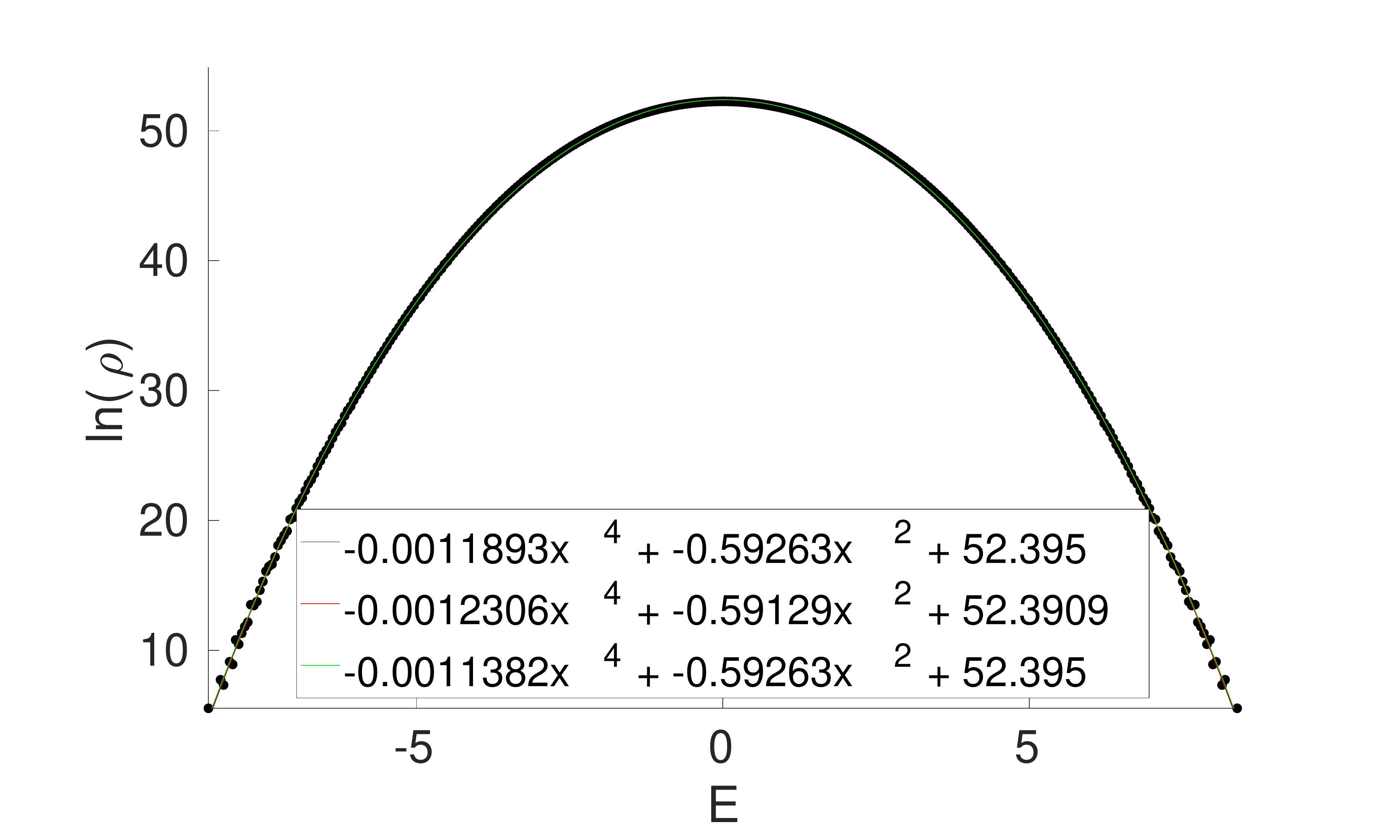}%
	\includegraphics[width=8.5cm]{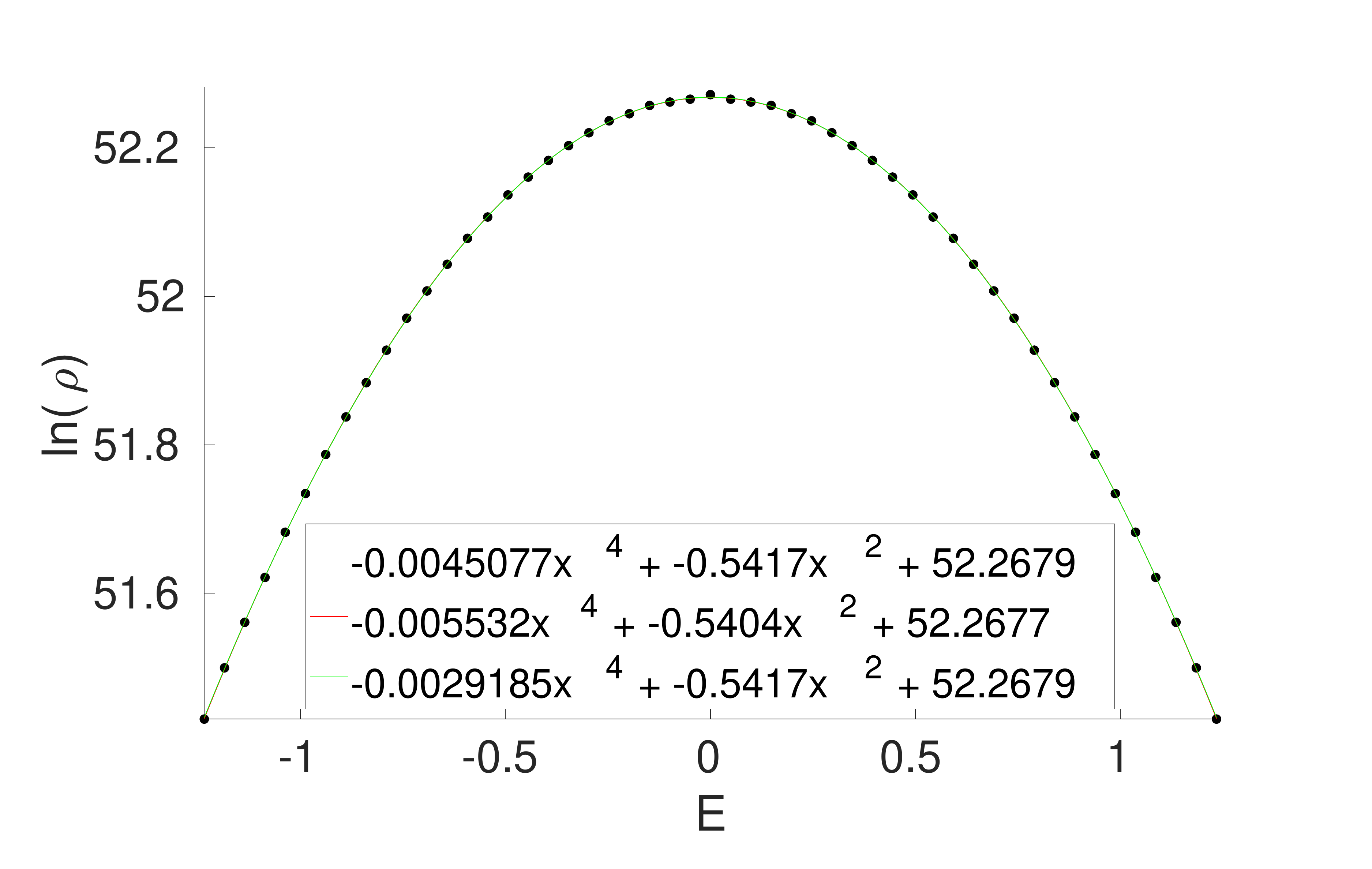}\\
	\caption{The spectrum for $N_1=N_2=9$. As one can see it has the same features as for $N_1=N_2=8$ and $N_1 = N_2 = 10$, but there is no separation between the even and the odd energy sectors. It could indicate that this difference has a purely group theoretic explanation.}
	
	\label{fig88}
	
\end{figure}

For $N_1= N_2=N$, we find that near $E\approx 0$ the density of states may be approximated by the Gaussian:
\begin{gather}
\log \rho(E) = N^2\log 2-\frac12 \left(\frac{ E}{\lambda N}\right)^2,
\end{gather}
where $\lambda=gN$ is the `t-Hooft coupling, which is held fixed as $N\to \infty$. We find nice agreement, which is shown for $N_1=N_2=8$ and $N_1=N_2=10$ in figure \ref{fig810} and for $N_1 = N_2 = 9$ in figure \ref{fig88}.

To demonstrate the validity of this approximation, let us compute 
\begin{gather}
\braket{E^n} = \int dE\, \rho(E)\, E^n = \frac{\tr\left[H^n\right]}{\tr\left[1\right]}\ .
\label{trafo}
\end{gather} 
This may be computed via the path integral
\begin{gather} \label{polycor}
\frac{\tr\left[H^n\right]}{\tr\left[1\right]} = \int \mathcal{D} \psi_{ab}\, H^n\, \exp\left(-\int \limits^\beta_0 d\tau\, \psi_{ab}(\tau)\, \partial_\tau \psi_{ab}(\tau) \right)\ .
\end{gather}
Therefore we can use standard Feynman techniques with the propagator $\braket{\psi_{ab}\psi_{a'b'}}=\frac12 \delta_{aa'} \delta_{bb'}$ and $H$ as an interaction vertex. Since $H$ has the form of a single-trace operator in the large $N$ limit, this product is dominated by the planar diagrams and moreover by the disconnected parts. From this point of view one can see that
\begin{gather}
\frac{\tr\left[H^{2n}\right]}{\tr\left[1\right]} = \frac{(2n)!}{2^n n!} \sigma_E^2,\quad \text{where}\quad  \sigma_E^2= \frac{\tr\left[H^2\right]}{\tr\left[1\right]} = 
\begin{tikzpicture}[baseline={([yshift=-.5ex]current bounding box.center)}]
\begin{scope}
\node at (-0.25,0) {$H$};
\node at (2.25,0) {$H$.};
\draw (0,0).. controls (1,1).. (2,0);
\draw (0,0).. controls (1,0.5).. (2,0);
\draw (0,0).. controls (1,-0.5).. (2,0);
\draw (0,0).. controls (1,-1).. (2,0);
\end{scope}
\end{tikzpicture}
\end{gather}
Then one can invert (\ref{trafo}) and get that $\rho(E)$ is the Gaussian distribution
\begin{gather}
\rho(E) = \frac{1}{\sqrt{2\pi \sigma_E^2}} \exp\left(-\frac{E^2}{2\sigma_E^2}\right)\ .
\end{gather}
The second moment, $\sigma_E^2$, is easy to compute using the diagrammatic technique: $\sigma_E^2 = g^2\left(N^4-N^3\right) \approx \left(\lambda N \right)^2$. To get the higher order corrections to the distribution function, we can continue calculating the energy moments, or we can instead simply compute the free energy and perform the inverse Laplace transformation to get the energy distribution. To be more precise, the free energy is defined as
\begin{gather}
F(\beta)=- \log \tr e^{-\beta H} =-\log \int dE\, \rho(E)\, e^{-\beta E}.
\end{gather}
This gives us a formula to compute $F(\beta)$ as a sum of the connected diagrams with $H$ as an interaction vertex
\begin{gather}
F(\beta) = \sum^\infty_{n=1} \beta^n \tr\left(H^n\right)_{\rm con} = \beta^2 \tr \left(H^2\right)_{\rm con} + \beta^4 \tr\left(H^n\right)_{\rm con} + \ldots
\end{gather}
Continuing this function to imaginary temperatures $\beta \to i\beta$, we can use the inverse Fourier transform
\begin{gather}
\rho(E) = \int \frac{d\beta}{2\pi} e^{i\beta E} e^{- F(i\beta)} = \int \frac{d\beta}{2\pi} e^{i\beta E } e^{-\beta^2 \tr \left(H^2\right)_{\rm con} - \beta^4 \tr\left(H^4\right)_{\rm con}+\ldots}.
\end{gather}
This integral can be calculated with the use of general diagrammatic technique, where $i E$ is the source for the energy, $\tr(H^2)_{\rm con}$ is the propagator, and  $\tr \left(H^4\right)_{\rm con}$ and the higher correlators are the vertices. By using these procedures we can compute the connected contribution. It is easy to compute the leading contributions to the connected trace of $H^4$,
\begin{gather}
\left(\tr{H^4}\right)_{\rm con.}=\left(\tr{H^4}\right) - 3 \left(\tr{H^2}\right)_{\rm con.}^2 = 8 g^4 N^6\ .
\end{gather}
After that we can restore
\begin{gather}
\log \rho(E) = N^2 \log 2 - \frac12 x^2 -\frac{1}{12 N^2} x^4 + \ldots,\quad E = g N^2 x\ .
\end{gather}
Comparing this expression with the numerical data we find a nice agreement between these two formulas.

\begin{figure}
	\centering
	\includegraphics[scale=1]{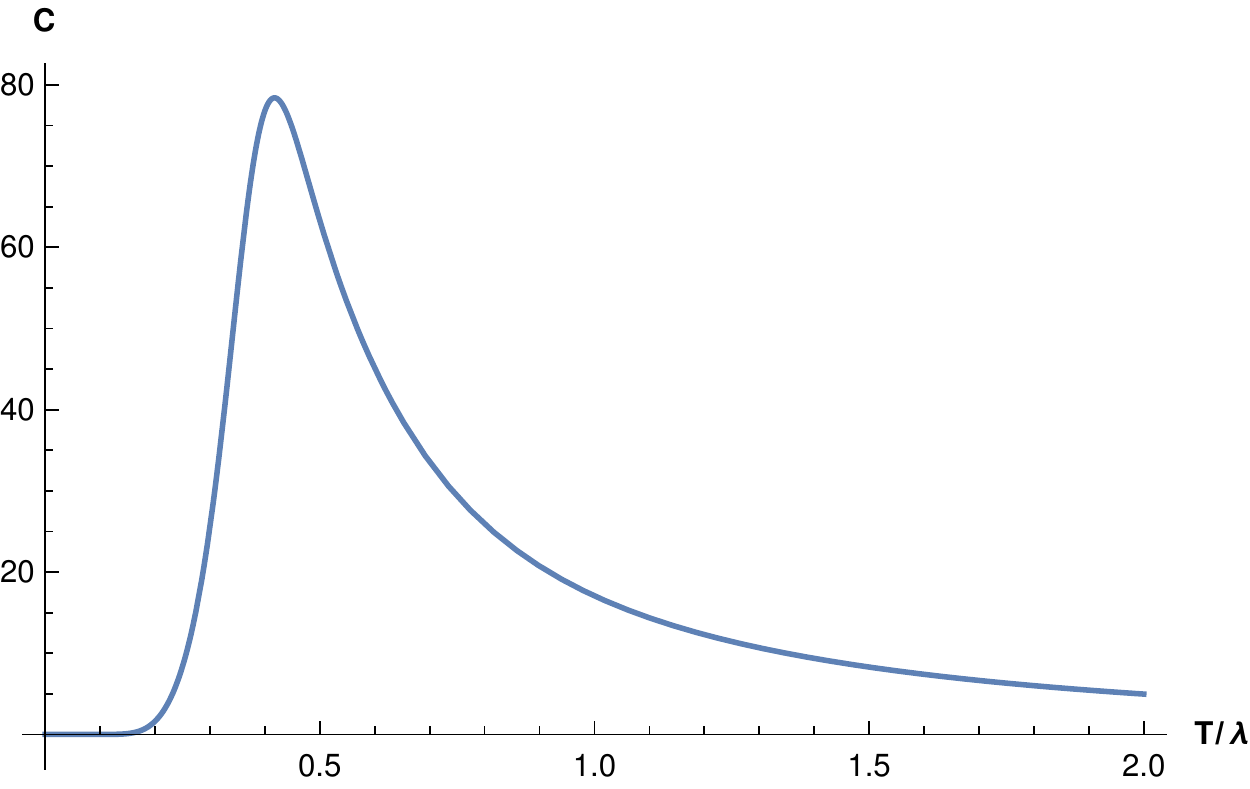}
	\caption{The specific heat as a function of temperature for the $O(N)^2\times O(2)$ matrix model with $N=10$. The low-temperature peak is due to the discreteness of the spectrum. At higher $T$, the specific heat falls off polynomially with the power $\alpha = \frac{d \log C}{d \log T} = -1.98$, close to that predicted by the analytic result \eqref{specdecrease}.}
	\label{fig:MatSpec}
\end{figure}
Let us note the splitting between the even and the odd energies, which is seen in figure \ref{fig810} but absent in figure \ref{fig88}.
These two sets of energies are distinguished by the value of 
\begin{gather}
P_C = \left(-1\right)^{\frac12\left(C^2_{O_1} - C^2_{O_2}\right)}.
\end{gather}
The trace of this operator counts the difference between the number of these branches. The trace of this operator over the whole space can be computed via the representation theory and is equal to $\tr P_C = 2^{2 N^2 - N + 1}$. 

We can study the thermodynamic properties of the matrix model in a similar fashion as in the case of the vector models. The behavior of the system would be analogous to a system of the spins in an external magnetic field. The partition function is
\begin{gather}
Z(T) = \int\limits^\infty_{-\infty} dE e^{- \frac{E}{T} } e^{-\frac{E^2}{2\lambda^2 N^2}} \sim e^\frac{\lambda^2 N^2}{2 T^2}, \quad F = - T \log Z(T) = -\frac{\lambda^2 N^2}{2 T},
\end{gather}
and the heat capacity $C$ is
\begin{gather}
C = - T \frac{\partial^2 F}{\partial T^2} = \frac{\lambda^2 N^2}{ T^2}\ .
\label{specdecrease}
\end{gather} 
This behavior is nicely captured by the numerical results shown in figure \ref{fig:MatSpec}. Note that the peak near $T_{\rm peak} \sim g \sim \frac{\lambda}{N}$ is due to the discreteness of the spectrum; it may be seen if we consider the contributions coming only from the ground state and the first excited state.

\section*{Acknowledgments}

This research was supported in part by the US NSF under Grants No.~PHY-1620059 and PHY-1914860. 
KP was also supported by the Swiss National Science Foundation through the Early Postdoc.Mobility Grant No. P2EZP2$\_$172168, and
 by DOE grant No. DE-SC0002140.
Some of the results presented here are from Gabriel Gaitan's Princeton University Senior Thesis (May 2019).
IRK is grateful to the Kavli Institute for Theoretical Physics at UC, Santa Barbara and the organizers of the program ``Chaos and Order: From strongly correlated systems to black holes" for the hospitality and stimulating atmosphere during some of his work on this project. His research at KITP was supported in part by
the National Science Foundation under Grant No. NSF PH-1748958.
We are grateful to S. Sondhi, P. Werner and C. Xu for useful discussions, and to A. Milekhin and G. Tarnopolsky for valuable discussions and comments on a draft of this paper. 
FKP is grateful to A.Morozov, D. Vasiliev,~S.Anokhina,~S. Shakirov,~A. Sleptsov,~K. Aleshkin and Y. Kononov for useful discussions and comments on the representation theory.
FKP is grateful to the Mainz Institute for Theoretical Physics (MITP) of the DFG Cluster
of Excellence PRISMA+ (Project ID 39083149), for its hospitality and its partial support.
PNP is grateful to the Yukawa Institute for Theoretical Physics (YITP) for its hospitality and partial support.

\appendix

\section{Decomposing the Hilbert Space}\label{HilSpace}

In this section we will review the structure of the Hilbert space of the $O(N_1)\times O(N_2)\times O(2)$ symmetric Majorana models.
We will study the irreducible representation of this algebra, which is spanned by $2\times N_1\times N_2$ Majorana fermions $\psi_{abc}$ subject to the anticommutation relations \eqref{HamAC}. To simplify the structure we introduce the Dirac fermions by combining two Majorana fermions,
\begin{gather}\label{complexferm}
c_{ab}=\frac{1}{\sqrt{2}}\left(\psi_{ab1}+i \psi_{ab2}\right),\, \bar{c}_{ab} = \frac{1}{\sqrt{2}}\left(\psi_{ab1}- i \psi_{ab2}\right),\, 
\notag\\
 \left\{c_{ab},\bar{c}_{a'b'}\right\}=\delta_{aa'}\delta_{bb'},\, \left\{c_{ab},c_{a'b'}\right\}=\left\{\bar c_{ab},\bar c_{a'b'}\right\}=0.
\end{gather}
These relations respect the larger symmetry group $U(N_1)_a\times U(N_2)_b$,  
and could be considered as symmetries of the Hilbert space, in contrast to the Hamiltonian \eqref{HamAC} which does not respect these symmetries. We can now try to decompose the Hilbert space in terms of the representations of these unitary groups using the character theory \cite{fulton2013representation}. We notice that the generator of the $U(N_1)_a$ and $U(N_2)_b$ groups could be rewritten in the following form
\begin{gather}
J^A_{T}= \frac{1}{2}T^A_{aa'} \left[\bar{c}_{ab}, c_{a'b}\right],\quad J^B_{T} =  \frac12 T^B_{bb'}\left[\bar{c}_{ab},c_{ab'}\right],
\end{gather}
where $T^{A,B}_{aa'}$ are hermitian matrices and can be considered as elements of the $\mathfrak{u}(N_i)$ algebra. Then the operators $J^{A,B}_T$ are the corresponding representations of these elements of the $\mathfrak{u}(N_i)$ algebra.
Hence, a general element of the $U_a(N_1) \times U_b(N_2)$ group, acting on the Hilbert space, is
\begin{gather}
g = e^{i T^A},\quad   \rho_\psi(g)=e^{ \frac{i}{2} T^A_{aa'} \left[\bar{c}_{ab}, c_{a'b} \right]}.
\end{gather}
Therefore we can compute the trace of this operator in the Hilbert space, and it is equal to the following:
\begin{gather}
\chi_{\mathcal{H}}(T^A,T^B) = \tr\left(e^{ \frac{i}{2}T^A_{aa'}\left[ \bar{c}_{ab}, c_{a'b}\right] + \frac{i}{2} T^B_{bb'} \left[\bar{c}_{ab}, c_{ab'}\right] }\right).
\end{gather}
We can study this trace rigorously and  expand this exponent to compute the trace order by order. Since the $T^{A,B}$ are hermitian matrices, we can diagonalize the matrix by some unitary transformation of the Hilbert space. Therefore, we can just consider the case where the matrices $T^{A,B}$ are diagonal
\begin{gather}
T^A_{aa'}= x_a \delta_{aa'},\quad T^B_{bb'}=y_b \delta_{bb'}.
\end{gather}
This gives the following formula for the character
\begin{gather}
\chi_{\mathcal{H}}(x_a,y_b) = \tr\left(e^{ \frac{i}{2} \sum_{a,b}(x_a+y_b)\left[ \bar{c}_{ab}, c_{ab}\right]}\right).
\end{gather}
Since each of the $N_1 N_2$ pairs $c_{ab},\bar{c}_{ab}$ and $\left[\bar{c}_{ab},c_{ab}\right]$ acts diagonally on the Hilbert space, the trace for each of the $ab$ effectively decouples from the rest making the computation straightforward,
\begin{gather}
\chi_{\mathcal{H}}\left(x_a,y_b\right) = \prod\limits^{N_1,N_2}_{a,b=1}\left(e^{-\frac{i}{2}(x_a+y_b)} + e^{\frac{i}{2}(x_a+y_b)}\right)  = \prod\limits^{N,M}_{a,b=1}2 \cos\left[\frac{x_a+y_b}{2}\right].
\end{gather}
One can see that this integral has the correct normalization, because if $x_a=y_b=0$ we restore the dimension of the space and $\chi_\mathcal{H} = 2 ^{N_1 N_2}$ as it should be. We can decompose this product in terms of the Schur polynomials, which are the characters of the irreducible representations of $U(N_i)$. Fortunately, this problem is easily solved with the use of the dual Cauchy identity \cite{macdonald1998symmetric}
\begin{gather}
 \prod\limits^{N_1,N_2}_{a,b=1}\left(e^{-\frac{i}{2}(x_a+y_b)} + e^{\frac{i}{2}(x_a+y_b)}\right) = \sum_{\lambda \subset (N_1^{N_2})} s_\lambda\left(e^{i x_a}\right) s_{\lambda^T}\left(e^{i y_b}\right),
\end{gather} 
where the $\lambda$ is the Young Tableaux and $\lambda^T$ is the transpose. Therefore the Hilbert space has a very simple decomposition in terms of the $U(N_i)$ groups. To each Young tableaux $\lambda \subset(N_1^{N_2})$ with no more than $N_1$ columns and $N_2$ rows we assign only one $U_a(N_1)$ representation; this state is an irreducible representation for the second unitary group described by the transposed Young Tableaux $\lambda^T$: $\mathcal{H} = \sum_{\lambda \subset (N_1^{N_2})} \left[\lambda\right] \otimes \left[\lambda^T\right]$.

Our original problem came from the study of the Hamiltonians and the anticommutation relations respecting the $O(N_i)$ group, instead of the unitary group $U(N_i)$. Since $O(N_i) \subset U(N_i)$ we can simply decompose each of the representations $[\lambda]$ of the $U(N_i)$ into irreducible representations of $O(N_i)$. This problem was successfully solved by Littlewood in 1947 \cite{littlewood1948invariant} and he obtained the following result \cite{koike1987young},
\begin{gather}
[\lambda]_{U(N_i)} = \sum_{\mu,\delta \prec \lambda, \delta \in \Delta_2} c^{\lambda}_{\delta, \mu} [\mu]_{O(N_i)},
\end{gather}
where $[\lambda]_{U(N_i)}$ and $[\mu]_{O(N_i)}$ are representations of the $U(N_i)$ and $O(N_i)$ groups described by Young Tableaux $\lambda$, and $\Delta_2$ is the set of all Young Tableaux with an even number of rows, and $c^{\lambda}_{\delta,\mu}$ is a Littlewood-Richardson coefficient. While this rule gives a nice procedure for the decomposition of the Hilbert space in terms of the irreducible representations of $O(N_i)$, it complicates the analytical understanding of the structure of the Hilbert space.

It is interesting to notice that if, instead of complex fermions $c_{ab}$, we considered Majorana fermions $\psi_{ab}$, we can compute the partition function to get the following character,
\begin{gather}
\chi_{\mathcal{H}}(x_a,y_b) = \prod^{\frac{N_1}{2},\frac{N_2}{2}}_{i=a,j=b}  \left(e^{i x_a}+e^{-i x_a}+e^{i y_b}+ e^{-i y_b}\right).
\end{gather}
We can deduce this structure heuristically. Note that, because of the Fermi-nature of each state $\lambda$ of the $O(2n)$ representation, we must include the correspondence $\lambda \subset ((N_1/2)^{(N_2/2)})$. One can compute the dimension of all of these representations and find that it is equal to the full Hilbert space. 
This gives a new dual Cauchy identity for orthogonal Schur polynomials,
\begin{gather}\label{CauchyOrth}
\sum_{\lambda \subset (n^m)} o_\lambda(x) o_{((N_1/2)^{(N_2/2)}/\lambda)'}(y) = \prod_{i,j} \left(x_i+x_i^{-1}+y_j+y_j^{-1}\right).
\end{gather}
It is easy to show that this is true just from the definition of the orthogonal characters. First of all, we notice that the charater of $O(2n)$ in the even case has the following form \cite{fulton2013representation,zhelobenko1962classical},
\begin{gather}
o_\lambda(x) =\frac{a_\lambda}{a_0}= \frac{\det\left(x_i^{\lambda_j+n-j} + x_i^{-(\lambda_j+n-j)}\right)}{\det\left(x_i^{n-j} + x_i^{-(n-j)}\right)}.
\end{gather}
Then we notice that if we denote the length of rows in the diagram $((N_1/2)^{(N_2/2)}/\lambda)'$ as $\mu_i$, the numbers $\mu_i+m-i,\lambda_j+n-j$ comprise a permutation $\sigma \in S_n$ of the numbers $I_n=\left\{0,1,\ldots,n+m-1\right\}$. Therefore, we just need to sum up all over possible permutations of the set $I_n$. This gives us
\begin{gather}
\sum_{\lambda \subset (n^m)} o_\lambda(x) o_{((N_1/2)^{(N_2/2)}/\lambda)'}(y) = \frac{\sum_{\sigma\in S_n}a_{\sigma{\lambda}}(x) a_{\tilde{\lambda}}(y)}{a_0^2},
\end{gather}
where $\sigma(\lambda) = \sigma\left(\{0,\ldots,n-1\}\right)$ $\sigma(\tilde{\lambda})=\sigma\left(\{n,\ldots,n+m-1\}\right)$. This could be rewritten using the Laplace rule for calculating determinants. We find that,
\begin{gather}
\sum_{\sigma\in S_n}a_{\sigma{\lambda}}(x) a_{\tilde{\lambda}}(y) = \Delta\left(x_1+x_1^{-1},x_2+x_2^{-1},\ldots, -y_1-y_1^{-1},\ldots, -y_n-y_n^{-1}\right) = \notag\\
=a_0(x) a_0(y) \prod^{n,m}_{i=1,j=1}\left(x_i+x_i^{-1}+y_j+y_j^{-1}\right) .
\end{gather}
The relation \eqref{CauchyOrth} directly follows. This concludes the proof of the structure of the $O(2n)\times O(2m)$ model. We can present a direct computation to show that this relation works for the $O(4)\times O(2)$ model. The content of the Hilbert space is
\begin{gather}
\mathcal{H} = \cdot \otimes \yng(2) + \yng(1) \otimes \yng(1) + \yng(1,1) \otimes \cdot
\end{gather}
The characters of this representations are
\begin{gather}
O(2):\chi_\cdot = 1, \quad \chi_{\yng(1)} = x_1 + x_1^{-1}, \quad \chi_{\yng(2)} = x_1^2 + x_1^{-2} \notag\\
O(4): \chi_\cdot =1 , \quad \chi_{\yng(1)} = y_1+y_1^{-1} + y_2+y_2^{-1},\quad \chi_{\yng(1,1)} =2+  y_1 y_2 + y_1 y_2^{-1} + y_1^{-1} y_2  + y_1^{-1} y_2^{-1}  .
\end{gather}
Substituting these into the character of the Hilbert space we get
\begin{gather}
\chi_H = \left(x_1+\frac{1}{x_1} + y_1 + \frac{1}{y_1}\right)\left(x_1+\frac{1}{x_1} + y_2 + \frac{1}{y_2}\right).
\end{gather}
As one can see, the representation of the one-dimensional fermions gives a very powerful tool for proving famous combinatorial equalities. It would be interesting to expand these ideas for other groups, say $Sp(N)$, and to generalize it for the case of MacDonald polynomials \cite{macdonald1998symmetric}.

\bibliographystyle{ssg}
%\bibliography{Spectrum2}
\bibliography{Hagedornv3}

\end{document}